\documentclass{article}

\usepackage{microtype}
\usepackage{graphicx}
\usepackage{subfigure}
\usepackage{booktabs} 
\usepackage{tcolorbox}

\usepackage{adjustbox,float}
\usepackage{array,multirow,graphicx}

\newcolumntype{R}[2]{%
    >{\adjustbox{angle=#1,lap=\width-(#2)}\bgroup}%
    l%
    <{\egroup}%
}
\newcommand*\rot{\multicolumn{1}{R{45}{1em}}}

\usepackage[hyphens]{url}

\usepackage{hyperref}

\usepackage[accepted]{2024arxiv}

\usepackage{amsmath}
\usepackage{amssymb}
\usepackage{mathtools}
\usepackage{amsthm}
\usepackage{graphicx}
\usepackage{xcolor}
\usepackage{balance}

\colorlet{darkgreen}{green!65!black}
\colorlet{darkblue}{blue!75!black}
\colorlet{darkred}{red!80!black}
\definecolor{lightblue}{HTML}{0071bc}
\definecolor{lightgreen}{HTML}{39b54a}
\definecolor{manyshot}{HTML}{6969ff}
\definecolor{medshot}{HTML}{f7c600}
\definecolor{fewshot}{HTML}{ff6969}
\definecolor{mypurple}{HTML}{412F8A}
\definecolor{myorange}{HTML}{fc8e62}
\usepackage{pifont}
\newcommand{\cmark}{\textcolor{darkgreen}{\ding{51}}}%

\usepackage[capitalize,noabbrev]{cleveref}

\theoremstyle{plain}

\theoremstyle{definition}

\theoremstyle{remark}

\usepackage[textsize=tiny]{todonotes}

\newcommand{\customtitle}{On the Standardization of Behavioral Use Clauses and Their Adoption for Responsible Licensing of AI}

\icmltitlerunning{\customtitle}
\begin{document}

\twocolumn[

\icmltitle{\customtitle}

\icmlsetsymbol{equal}{*}

\begin{icmlauthorlist}
\icmlauthor{Daniel McDuff}{}
\icmlauthor{Tim Korjakow}{}
\icmlauthor{Scott Cambo}{}
\icmlauthor{Jesse Josua Benjamin}{}
\icmlauthor{Jenny Lee}{}
\icmlauthor{Yacine Jernite}{}
\icmlauthor{Carlos Muñoz Ferrandis}{}
\icmlauthor{Aaron Gokaslan}{}
\icmlauthor{Alek Tarkowski}{}
\icmlauthor{Joseph Lindley}{}
\icmlauthor{A. Feder Cooper}{}
\icmlauthor{Danish Contractor}{}

\end{icmlauthorlist}

\icmlcorrespondingauthor{Daniel McDuff}{dmcduff@uw.edu}
\icmlcorrespondingauthor{Danish Contractor}{danish.contractor.007@gmail.com}

\icmlkeywords{Responsible, AI, Licensing}

\vskip 0.3in
]

\begin{abstract}
Growing concerns over negligent or malicious uses of AI have increased the appetite for tools that help manage the risks of the technology. In 2018, licenses with behaviorial-use clauses (commonly referred to as Responsible AI Licenses) were proposed to give developers a framework for releasing AI assets while specifying their users to mitigate negative applications. As of the end of 2023, on the order of 40,000 software and model repositories have adopted responsible AI licenses licenses. Notable models licensed with behavioral use clauses include BLOOM (language) and LLaMA2 (language), Stable Diffusion (image), and GRID (robotics). This paper explores why and how these licenses have been \textit{adopted}, and why and how they have been \textit{adapted} to fit particular use cases. We use a mixed-methods methodology of qualitative interviews, clustering of license clauses, and quantitative analysis of license adoption. Based on this evidence we take the position that responsible AI licenses need \textit{standardization} to avoid confusing users or diluting their impact. At the same time, \textit{customization} of behavioral restrictions is also appropriate in some contexts (e.g., medical domains). We advocate for ``standardized customization'' that can meet users' needs and can be supported via tooling.\looseness=-1  
\end{abstract}

\section{Introduction}
Openness is a tenet of scientific research and plays an important role in the development of new technologies.
By making assets available to third parties, scientific results can be verified, and systems can be interrogated, tested and audited~\citep{resnik2006open, vonkrogh2006open}. 
In AI, significant advances have been made possible thanks to the open sharing of data, code, models, and applications~\citep{gokaslan2019openwebtext, gokaslan2023commoncanvas, workshop2022bloom, raffel2020t5, rombach2022high, touvron2023llama}. 
The fact that researchers and developers can use, modify or extend what others have built enables an important form of AI decentralization and supports accessibility.
\looseness=-1 

However, openness has also come with significant tensions, particularly for pretrained (so-called \emph{base} or \emph{foundation}) models. On the one hand, these models can be used in a variety of domains, often with little or no finetuning~\citep{radford2019language,brown2020language}. On the other, this versatility means that they can be used by different actors in ways that are not aligned with the applications intended by their creators~\citep{lee2023talkin}. Some of these uses may be overtly harmful (e.g., generating content to deceive or harass a person/people) and others may present unintended higher risks (e.g., generating diagnoses that may sometimes be inaccurate or biased, extracting PII from training data)~\citep{epic, nasr2023scalable, milesblog}.
Combined with decentralization, such downstream uses can present challenges for accountability and recourse~\citep{cooper2022accountability, cooper2023report}.\looseness=-1
\begin{figure*}[t]
  \centering
  \includegraphics[width=\linewidth]{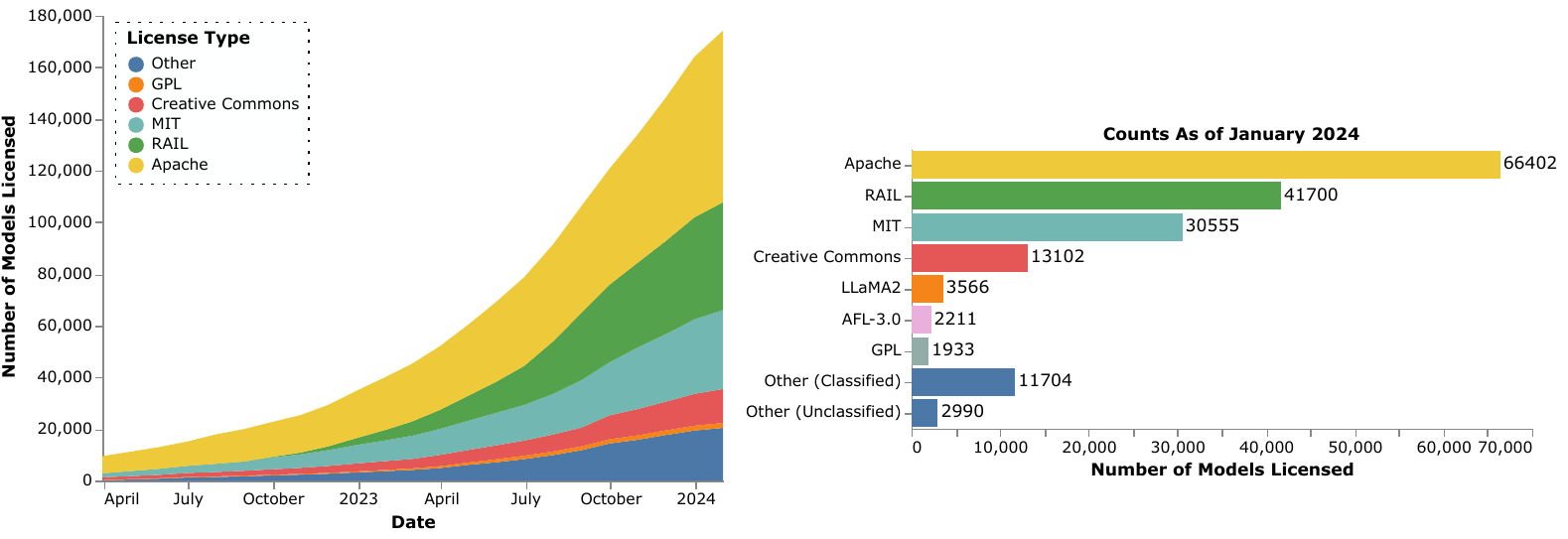}
  \caption{\textbf{Adoption of Licenses with Behavioral Use Clauses.} The number of repositories by license type on the HuggingFace model hub. As of January 2024, 41,700 RAIL licensed repositories and 3,566 LLaMA2 licensed repositories existed, both of these licenses include behavioral-use clauses.}
  \label{fig:quant_results}
\end{figure*}

\textbf{Usage restrictions on contractual agreements.} 
The release of assets by private organizations is usually accompanied by a provider-user contractual agreement. Users must agree to terms specified by the provider for unhindered and continued access to the service. While traditionally such terms were geared towards compliance with laws, mitigating legal risks and protecting intellectual property, some AI providers have begun to include additional clauses that govern usage. For example, Open AI's policies disallow generation of content for dissemination in electoral campaigns~\citep{openaiusage},  Microsoft's FaceAPI services are also subject to ``specific license terms'' and are only available in limited access to ``customers managed by Microsoft''~\citep{microsoftpolicy}. 

\textbf{To release or not to release.}  Not all researchers or research teams have the resources to create customized legal agreements for the AI models or source code they would like to release. The choice often becomes to release the code with no restrictions or to not release it at all. This forces the creator to decide between openness and democratization on one hand, and responsible use on the other.

While tools for software licensing do exist~\cite{bretthauer2001open}, historically the most frequently adopted licenses (e.g., MIT, Apache) do not contain restrictions on how code, models or applications are to be \emph{used}. Adapting licenses as a tool for responsibly releasing AI software was proposed by Contractor et al.~\cite{contractor2022behavioral}. These licenses contain \emph{behavioral use clauses} that enable software and models to be released with restrictions around how they are \emph{used}. The paper, proposed that these licenses could be implemented to complement existing responsible AI guidelines.\looseness=-1

\textbf{Licenses with Behaviorial-use Clauses.} Over the past five years, licenses with behavioral use clauses (BUC) have been gaining adoption at an increasely rapid rate (see Fig.~\ref{fig:quant_results}). In this paper, we refer to responsible AI licenses as a broad category of licenses that incorporate BUC.\footnote{Licenses using the acronym `RAIL' are specific variant of the broader class of responsible AI licenses.} 
Licenses that allow adaptation and reuse of software, models, data, or applications can be made dependent on behavioral use conditions.
However, each different type of asset has their own ideosyncracies. 

A study by OpenFutures of 39,000 repositories found a clear trend towards the adoption of responsible AI licenses~\cite{Keller2023Growth}. OpenRAIL licenses~\citep{openrail}, a specific variant of RAIL licenses, were the second most used license category. To date, such licenses have primarily been applied to AI models. 

For example, BLOOM~\cite{workshop2022bloom} is a large parameter multilingual language model, and accompanying BigScience OpenRAIL license enables derivative uses but restricts applications amongst others that violate laws, generate or disseminate verifiably false information, or predict if an individual will commit fraud/crime.  Stable diffusion~\cite{rombach2022high} ``provided weights under a license to prevent misuse and harm as informed by the model card~\cite{mitchell2019model}, but otherwise remains permissive.''   Subsequently the LLaMA 2~\cite{touvron2023llama} and FALCON~\cite{almazrouei2023falcon} models were released with intersecting behavioral use restrictions, yet FALCON has a smaller set of clauses than the others.

The use of responsible AI licenses is not restricted only to foundation models, Robotics platforms (GRID)~\cite{vemprala2023grid}, edge IoT systems~\cite{janapa2023edge} and medical sensors~\cite{liu2023rppg} that use AI components have also adopted similar clauses.  In the realm of training data, similar approaches to licensing have also been the subject of discussion \cite{liDataLicense} and experimentation. For instance, AI2 created the ImpACT license to apply broadly to ML artifacts that include both models and data \cite{AI2ImpACTLicenses} and used an ImpACT license for the DOLMA dataset release \cite{soldaini2023dolma}.

\textbf{Contributions.} While the growth of interest in, and adoption of, these licenses is very apparant, there is no single standard license. In this paper we explore the reasons for the trend towards behavioral use licensing, the proliferation of different licenses and choice of clauses, and the need for standardization. We perform a quantitative analysis of the licenses used in over 170,000 model repositories and highlight the growing trend toward adopting responsible AI licenses. We also qualtitatively evaluate the similarities and differences between the specific license clauses included in these artifacts. We then report on semi-structured interviews conducted with researchers who have released high-profile AI models and software with responsible AI licenses (across computer vision, natural language, 
and robotics).\textbf{We take the position that responsible AI licenses need \textit{standardization} to avoid confusing users or diluting their impact. At the same time, \textit{customization} of behavioral restrictions is necessary and appropriate in some contexts (e.g., medical domains) and can be supported by tooling.}

\section{Regulation and Licensing for AI}

Regulation and licensing of AI systems is difficult because of their complex supply chains and the uncertainty about how copyright rules apply to AI systems~\cite{lee2023talkin}. One must consider the licensing of the data, the code, the models, and the machine learning libraries used to train the models. New considerations are being raised as the technology is developed; for example, courts are now examining the relationship between the licensing of the data and the models where they have not in the past. Outcome-based regulation of relatively simple classifiers has proven difficult~\cite{cooper2022accountability}, let alone regulation for large foundation models. Some academics believe that training AI models on copyrighted data is fair use under current US copyright law~\citep{lemley2021fairlearning, fairuse}, while others argue that there is likely not a blanket fair-use rationale for all possible systems and their possible outputs~\citep{lee2023talkin, samuelson, sag2023safety}. 
Courts and legislators, if asked existential questions about whether large-scale AI models should exist, may realistically decide that using copyrighted data to train AI models does not constitute copyright infringement, and that the fair use doctrine may be invoked to shield such use.\footnote{However, there are many copyright issues beyond fair use and copyrighted training data. See~\citet{lee2023talkin}.}

\noindent {\bf How can regulation help?}:
Regardless of the currently-unresolved intellectual property regime governing AI systems, the aim of regulation is to provide an enforceable mechanism to govern how AI is or is not used -- for example, by focusing on data privacy. Many countries, states, and cities are using regulations to avoid undesired uses of AI. For instance, the city of San Francisco banned the use facial regulation in CCTV cameras, and imposed a series of certifications for the use of self-driving cars in its streets. In Europe, the General Data Protection Regulation GDPR\footnote{\url{https://gdpr.eu/}} requires data collectors to state the purpose for which data is being collected, to obtain consent, and then to collect only the minimum amount of data required to accomplish the task for which consent was received. Additionally, it imposes requirements on data processors to be able to verifiably remove data when consent is revoked. 

The recently proposed EU AI Act,\footnote{\url{https://digital-strategy.ec.europa.eu/en/policies/regulatory-framework-ai}} bans certain uses of AI (like the use of biometric identification in public spaces by governments) and imposes conditions (such as conformity assessment, trustworthy AI properties, and human oversight) for the use of AI in certain domains considered high-risk, such as hiring and education. However, solely relying on regulations to prevent undesired applications of AI presents several difficulties. 

First, in contrast to other fields of science, AI is moving very rapidly and there is a lack of specific regulation at the present time. Significantly more advanced models and systems are being released every month. The introduction of transformers~\cite{vaswani2017attention}, coupled with the increased availability of computing resources, has resulted in rapid advancement in the areas of natural language generation and multi-modal reasoning, in under five years. We argue that while government regulation processes are best equipped to handle the effect of a new technology in the long term, they leave a gap in terms of governance in the short term.

Second, often a responsible approach to releasing AI systems needs to consider the capabilities and limitations of a specific AI model. For instance, if a machine translation system has a high word-error rate, it may not be a problem if used for recreational purposes, but could be dangerous when employed in other contexts (e.g., translating documents with medical or legal implications). Such case-specific determinations can be difficult to encode in a legislation that aims to specify general rules to balance different broad categories of risks, and even harder to operationalize in the absence of meaningful and generally applicable thresholds. 
Further, different parties, developers or researchers may want to impose specific restrictions on the use of the AI models they build in cases where they feel government regulation is too weak or does not exist.

\noindent {\bf Can licenses help?}
Recognizing these challenges, the use of contractual and copyright law have been recently suggested as a means of controlling end-use~\cite{contractor2022behavioral}. This approach holds promise, after all much of the software in use today is accompanied by a license governing its terms-of-use. Such terms-of-use can govern commercial use, compliance with local laws and regulation, dispute resolution mechanisms, disclaimers of warranties and risks and intended-use. This is not only true for software sold by companies, source code that is open-sourced is accompanied by a license which grants the permissions for licensee to distribute, re-use, modify the accompanying software and source code. In some cases, terms referred to as ``copy-left'' clauses are included in some open-source licenses, which require that any downstream software or code reusing any component of the licensed artifact is subject to the same requirements as the original license. 
We argue that licensing will become increasingly important as AI continues to develop at a rapid pace.

\textbf{Related Tools for AI Governance.}
A number of other mechanisms aimed at promoting the ethical and responsible use of AI have been proposed that are relevant to mention. 

\noindent {\textit{Promoting Informed-use:}} A number of tools have been proposed as a way to organize infrormation about machine learning models in a consistent manner. AI factsheets~\cite{FactSheets} and AI model cards~\cite{ModelCards} both help to achieve these ends.  However, they themselves do not have an explicit mechanism for enforcement about how the models are used.

\noindent {\textit{Ethical Guidelines:}}
Ethical guidelines\footnote{\url{https://www.microsoft.com/en-us/ai/responsible-ai-resources}} \footnote{\url{https://ai.google/responsibilities/responsible-ai-practices/}} and Responsible AI initiatives\footnote{\url{https://pwc.com/rai}} \footnote{\url{https://www.acm.org/binaries/content/assets/public-policy/final-joint-ai-statement-update.pdf}} help to communicate best practices and standards that can help practioners act responsibly. Some guidelines cover specific use cases, but others are only about general practicies.

\section{A Study of AI Licenses}

\subsection{Methodology}

To motivate our position on how responsible AI licenses should evolve, we conducted a study of the current state of the licensing in AI. Given the complex socio-technical nature of of the domain we pursue a mixed-methods approach combining insights from interviews of licenses users, quantitative analysis of license adoption and a review of existing licenses including clustering of their behavioral use clauses.
The interviews involved practitioners and subject matter experts who have adopted licenses for high profile AI projects. We analyzed over 170,000 machine learning model repositories. To cluster the licenses we manually categorized behavioral use clauses in the most heavily utilized license types. 


\subsection{Interviews with License Adopters}
\begin{figure}[t]
\begin{tcolorbox}
\small
\textbf{Q1.}  Describe your background and the project that you chose to license with behavioral use clauses. 

\textbf{Q2.} Why did you choose to release the model with behavioral use clauses?

\textbf{Q3.} Did you feel it was necessary to modify an existing responsible AI license? Why/why not?

\textbf{Q4.} What considerations drove your choices of clauses to include in the license?

\textbf{Q5.} What other restrictions did you include in the final license?

\textbf{Q6.} What other restrictions did you consider including in the license, but ultimately excluded?

\textbf{Q7.} What do you think the restrictions will achieve in your case?

\textbf{Q8.} How can use restriction based licenses have impact in the AI sector?

\textbf{Q9.} Imagine: What tools, platforms or processes could accompany use restriction based licenses to increase their efficacy, adoption or robustness?
\end{tcolorbox}
\caption{Questions posed to our interviewees regarding the adoption of responsible AI licenses.}
\label{fig:questions}
\vspace{-5ex}
\end{figure}
\textbf{Participants.} We performed semi-structured 30-minute interviews with four AI researchers and software developers. The interviewees has a mean of 18 years of post-graduate experience in the field of AI (8, 30, 10, 24 years). The interviews explored the questions listed in Fig. \ref{fig:questions}.  The interviewees had all release projects in the past two years under a responsible AI license.  These included: large language and code models, and robotics and computer vision software. 

\begin{figure*}[t]
  \centering
  \includegraphics[scale=0.6]{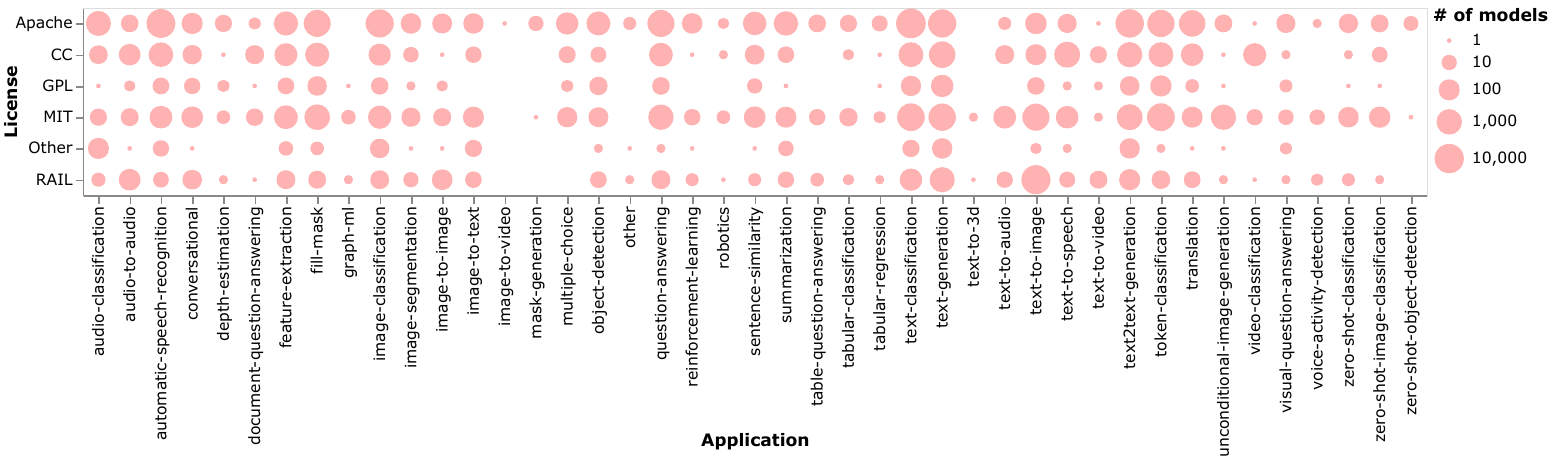}
    \caption{\textbf{Applications of Models by License Type.} The number of repos in each application domain that are under each license.}
  \label{fig:applications}
\end{figure*}

\textbf{Analysis.} A common theme the interviewees discussed was the tension between the ``willingness to release assets on an open basis'' on one hand and the ``concerns within the community and internal values about uses of the model.'' (P1).  This was a motivating factor for the choice to select a responsible AI license rather than a traditional OS license. With the evolving nature of AI there is a ``need for people to think about how to use [models] responsibly'' (P2).  

The interviewees spoke about the evolving nature of AI technology as one of the reasons they chose to adapt use clauses in the licenses. 
One interviewer summarized a common theme describing that ``existing OS licenses didn't perfectly match our mission'' (P3). 
One roboticist, commented on the fact that physical instantiations of AI required terms that restricted ``destruction of physical people of property'' that other licenses did not include. A second point was that existing OS licenses were ``not suitable for [other assets]'' (e.g., datasets) (P3).

In the process of considering their options, a practice that multiple interviewees used was to review existing OS licenses to see if any satisfied their needs. However, licenses like Apache 2.0 were not ``ML or LLM specific'' (P1) and needed to be adapted. They noted that adaptation was not trivial. One participant, who was an ML researcher and entrepenur, said licensing was ``not our expertise, and would have required resources that we didn’t have'' (P2). Another commented on the ambiquity introduced by changing licenses ad hoc - ``[It is difficult] for someone to know the difference between license A and B, that is why we didn’t modify an OS license'' (P4). When asked about tooling, the participants identified that there were other situations in which agreements could be ``generated via questionnaire ... by ticking boxes'' and that having a template would be easier.  

\subsection{Analysis of Licenses and their Clauses}

\textbf{Quantitative Analysis.} We performed an analysis of the \href{huggingface.co/docs/hub/models-the-hub}{HuggingFace Model Hub}. The Model Hub API provides information about the models, licenses and domains/applications, when the repository was created and how many times the model has been downloaded. At the time of the analysis (January 2024) 174,163 model repositories had licenses associated with them.

Of these, the most common licenses used were Apache\footnote{\url{https://www.apache.org/licenses/LICENSE-2.0}} (38.1\%), RAIL\footnote{\url{https://www.licenses.ai}} (24.0\%), MIT\footnote{\url{https://opensource.org/license/mit/}} (17.5\%) and Creative Commons\footnote{\url{https://creativecommons.org/licenses/list.en}} (7.5\%) (see Fig.~\ref{fig:quant_results}).
RAIL licenses are the most common type of license \emph{with} behavioral use clauses and were the second most adopted license type overall (see Fig.~\ref{fig:quant_results}). The share of RAIL licenses has grown from 1\% (N=209) in September 2022 to 10\% (N=4,035) in January 2023 to 24.0\% (N=41,700) in January 2024. 
In comparison to counting the number of repositories licensed under RAIL licenses, RAIL licenses measured by the number of downloads make up the fourth biggest category behind Apache, MIT and CC licenses (see Fig. \ref{fig:treemap} in the appendix). These results support the conclusion that RAIL adoption is considerable and growing; however, some of the most popular models are not using RAIL licenses.

RAIL licenses have been used with repositories across many different application domains, including text, image, speech processing and many methods, including supervised and unsupervised learning, reinforcement learning, graph learning, classification and generative methods. Given the adoption by StableDiffusion~\cite{rombach2022high}, text-to-image translation was the most common model type to adopt a behavioral use license (see Fig.~\ref{fig:applications}).

\begin{table*}[ht]
	\centering
	\scriptsize
	\setlength\tabcolsep{6pt} 
 \begin{tabular}{cp{9cm}ccccccc}
	& \textbf{Behavioral Restrictions} & \rot{\textbf{AIPubs RAIL}} & \rot{\textbf{BigSci. OpenRAIL}} & \rot{\textbf{CodeML OpenRAIL}} & \rot{\textbf{LLaMA 2}} & \rot{\textbf{FALCON}} & \rot{\textbf{ImpACT L/M/H}} & \rot{\textbf{GRID}}  \\ 
 \toprule 
      \parbox[t]{2mm}{\multirow{5}{*}{\rotatebox[origin=c]{90}{\textbf{Discrim.}}}} &
    (1) Discriminate against people based on legally protected characteristics & \cmark & \cmark & \cmark & \cmark & & & \cmark \\
    & (2) Administrate justice, law enforcement, immigration, or asylum processes, such as predicting that a natural person will commit a crime. &  & \cmark & \cmark & & & \cmark & \cmark \\
    & (3) Defame, disparage or otherwise harass others. & \cmark & \cmark & \cmark & \cmark & \cmark & & \cmark \\
    & (4) Exploiting, harming or attempting to exploit or harm minors in any
way & \cmark & \cmark & \cmark & \cmark & \cmark & & \cmark \\
   \toprule
   \parbox[t]{2mm}{\multirow{5}{*}{\rotatebox[origin=c]{90}{\textbf{Disinformation}}}}  & (5) Create, present or disseminate verifiably false or misleading information for economic gain, harm people, or to intentionally deceive the public. & \cmark & \cmark & \cmark & \cmark & & & \cmark \\
    & (6) Synthesize or modify a natural person’s appearance, voice, or other individual characteristics, unless prior informed consent of said person is obtained. & \cmark & \cmark & \cmark & \cmark & & & \cmark \\
    & (7) Generate/disseminate information and place the information in any public context without expressly disclaiming that the content is machine generated. &  \cmark & \cmark & \cmark & \cmark & & \cmark & \cmark \\
    \toprule
    \parbox[t]{2mm}{\multirow{2}{*}{\rotatebox[origin=c]{90}{\textbf{Legal}}}} & (8) Fully automate decision-making that creates, modifies or terminates a binding, enforceable obligation between entities. & \cmark & \cmark & \cmark & & & \cmark & \cmark \\
    & (9) Violate any national, federal, state, local or international law or regulations. & \cmark & \cmark & \cmark & \cmark & \cmark & & \cmark \\
    \toprule
    \parbox[t]{2.5mm}{\multirow{3}{*}{\rotatebox[origin=c]{90}{\textbf{Privacy}}}} & (10) Utilize personal information to infer additional personal information, including legally protected characteristics, vulnerabilities or categories; unless informed consent from the data subject is received. & & & & \cmark \\
    & (11) Generate/disseminate personal identifiable information that can be used to harm an individual or to invade the personal privacy of an individual.  & \cmark  & \cmark  & \cmark & \cmark & & & \cmark \\
    & (12) Engage in, promote, incite, or facilitate the harassment, abuse, threatening, or bullying of individuals or a group.  & \cmark & \cmark & \cmark & \cmark & \cmark & & \cmark \\
    \toprule
    \parbox[t]{2mm}{\multirow{4}{*}{\rotatebox[origin=c]{90}{\textbf{Health}}}} & (13) Provide medical advice or clinical decisions without accreditation of the system; unless the use is (i) in an internal research context with independent and accountable oversight and/or (ii) with medical professional oversight. & & \cmark & \cmark & & \cmark \\
    & (14) In connection with any activities that present a risk of death or bodily harm to individuals, including self-harm or harm to others, or in connection with regulated or controlled substances.  & & & & \cmark\\
    & (15) In connection with activities that present a risk of death or bodily harm to individuals, including inciting or promoting violence, abuse, or any infliction of bodily harm. & & & & \cmark \\
    & (16) Exploitaion of vulnerabilities of a group/person in a manner that is likely to cause physical or psychological harm & \cmark & \cmark & \cmark \\

    \toprule
    \parbox[t]{2mm}{\multirow{4}{*}{\rotatebox[origin=c]{90}{\textbf{Military}}}} & (17) Use for active deployment of weaponry. & & & & & & & \cmark \\
    & (18) Use for purposes of building or optimizing military weapons or in the service of nuclear proliferation or nuclear weapons technology. & & & &\cmark & & \cmark & \\
    & (19) Use for purposes of military surveillance, including any research related to military surveillance. & & &  &\cmark & & \cmark \\
    & (20) ‘Real time’ remote biometric processing or identification systems in publicly accessible spaces for the purpose of law enforcement. & & & & & & \cmark & \cmark \\
    \toprule
    \parbox[t]{2mm}{\multirow{2}{*}{\rotatebox[origin=c]{90}{\textbf{Other}}}} & (21) Generate/disseminate malware/ransomware or other content for the purpose of harming electronic systems. & & & \cmark \\
    & (22) Fail to appropriately disclose to end users any known dangers of your AI system & & & & \cmark \\
    \bottomrule
  \end{tabular} 
 \caption{\textbf{Summary of Behavioral-Use Clauses.} Clauses included in popular responsible AI licenses.}
	\label{tab:methods}
\end{table*}

\begin{table*}[ht]
\scriptsize
\centering
\begin{tabular}{rccccccc}
& \multicolumn{1}{c}{\textbf{\begin{tabular}[c]{@{}c@{}}AI Pubs\\  RAIL\end{tabular}}} & \multicolumn{1}{c}{\textbf{\begin{tabular}[c]{@{}c@{}}BigScience\\  OpenRAIL\end{tabular}}} & \multicolumn{1}{c}{\textbf{\begin{tabular}[c]{@{}c@{}}CodeML \\ OpenRAIL\end{tabular}}} & \multicolumn{1}{c}{\textbf{Llama 2}} & \multicolumn{1}{c}{\textbf{FALCON}} & \multicolumn{1}{c}{\textbf{\begin{tabular}[c]{@{}c@{}}ImpACT L/M/H\\ Licenses\end{tabular}}} & \multicolumn{1}{c}{\textbf{GRID}} \\ \toprule
\textbf{Behaviorial-Use Restrictions}  & \cmark                                                            & \cmark                                                                   & \cmark                                                               & \cmark            & \cmark           & \cmark                                                                    & \cmark         \\
\textbf{Conditions on Derivatives}                &                                                                                      &                                                                                             &                                                                                         & \cmark            &                                     &  For H only                                                                                            &                                   \\
\textbf{Conditions on Distribution}    & \cmark                                                            &                                                                                             &                                                                                         &                                      &                                     & For M \& H only                                                                              &                                   \\
\bottomrule                              
\end{tabular}
\caption{\textbf{Requirements for Derivatives and Distributions.} All licenses require inclusion of the original behavioiral-use restrictions; this table reports the presence of additional conditions imposed on derivatives and distributions.} \label{tab:conditions}
\end{table*}

\textbf{Qualitative Analysis.} 
To better understand the types of behavioral uses that have been restricted in the first wave of responsible AI licenses we conducted a qualitative analysis of $77$ different legal clauses from $7$ different responsible AI licenses that are currently in use:  AI Pubs RAIL\footnote{\url{https://www.licenses.ai/blog/2023/3/3/ai-pubs-rail-licenses}} -- which is also available for use by researchers publishing at AAAI conference venues,\footnote{\url{https://aaai.org/aaai-publications/}} BigScience OpenRAIL~\cite{workshop2022bloom}, CodeML OpenRAIL~\cite{li2023starcoder}, Llama-2~\cite{touvron2023llama}, FALCON~\cite{almazrouei2023falcon}, ImpACT Licenses~\cite{AI2ImpACTLicenses} and GRID~\cite{vemprala2023grid}.

We guided our analysis with the following questions: Q1: What artifacts are being restricted?, Q2: What are the uses being restricted?, and Q3: How are they being restricted? For Q1, we coded for the specific language referencing the artifact whose use is being restricted in the clause. For Q2, two of the authors independently coded broadly for any concepts that could help us understand the values and social norms of the developer community. Each author independently conducted an axial coding phase \cite{braunclark06} to refine the granularity of our analysis. 
Next, we mapped the resulting codes to existing Responsible AI frameworks like the NIST AI RMF \cite{NISTRMF2023} and the CSET Harm Taxonomy \cite{CSETAI}. Each of these frameworks are designed to 
facilitate a more standardized and informed, yet flexible and adaptable approach for measuring, monitoring, and mitigating AI harm. For Q$3$, we learned early on that nearly all restrictions in the clauses we analyzed restrict a particular use of an artifact by simply prohibiting it. While the results of the question as initially posed would not be interesting with our data, we did observe that some use restrictions had caveats.

The result was 22 clauses (see Table~\ref{tab:methods}). We find that almost all licenses include clauses that govern discrimination, disinformation and violations of the law. Most contain privacy related restriction. The other two main groups of clauses are heathcare and military applications of which some licenses include restrictions.

\textbf{Additional Restrictions on Distribution.} Apart from behavioral-use clauses applicable to the use and distribution of the artifacts, licenses sometimes also include additional requirements (See table \ref{tab:conditions}). For instance, the AI Pubs RAIL License permits the use of the licensed code or model only for ``Research-use'' while the LLaMA-2 license explicitly prohibits the use of the model for anything except creating a derivative of LLaMA-2 with the same terms. It also stipulates additional licensing requirements if the monthly active users of the model exceed 700 million. The ImpACT licenses permit the creation of derivatives and allow distribution for low-risk~(L) use-cases but do not allow derivatives for high-risk~(H) use-cases and they also do not permit derivatives for both, medium~(M) and high-risk~(H) use-cases. 

The flavor of RAIL licenses, known as the OpenRAIL family of licenses, avoid non-behavioral use restrictions (such as ``research only'' requirements). Currently, the ``OpenRAIL'' moniker is given to license terms that (1) incorporate behavioral-use clauses, and (2) allow for the otherwise unlimited use, modification, and distribution of applicable artifacts, as long as the behavioral-use restrictions are included. The ``open'' designation has been important to certain actors within the AI eco-system who wish to emphasize that the IP they are sharing can be treated much like other open source code (freely share-able and modifiable), albeit with behavioral-use restrictions. Both the BigScience and CodeML RAIL Licenses are OpenRAIL Licenses. 

\section{Discussion}

\textbf{Responsible AI Licenses Are Being  \underline{Adopted}.} Our quantitative analysis of the repositories shows that RAIL licenses are being adopted at a relatively large scale. This has been driven by a number of high profile and widely adopted projects and the associated derivative projects. The adoption is large enough to suggest that it is necessary to seriously study these types of licenses and their implications for the AI and open source communities, the latter of whom have been grappling with the definition of Open Source AI.\footnote{\url{https://deepdive.opensource.org/wp-content/uploads/2023/02/Deep-Dive-AI-final-report.pdf}}  

The rapid rise of licenses with behavioral use clauses, particularly amongst projects involving large parameter models, signals a perception that existing open source licenses do not contain sufficient restrictions for responsibly distributing capable machine learning models. All of our interviewees highlighted that conundrum.

\textbf{Responsible AI Licenses Are Being  \underline{Adapted}.}
Our analysis of behavioral use clauses included in responsible AI licenses shows that there is a large amount of overlap in restrictions.  However, they differ in important ways. Some licenses (e.g., FALCON) are more permissive and include a smaller subset of clauses while others (e.g., OpenRAIL, ImpACT) include a much larger list. Some licenses (e.g. CodeML OpenRAIL) include domain specific restrictions.

\begin{figure*}
    \centering
    \includegraphics[width=1\textwidth]{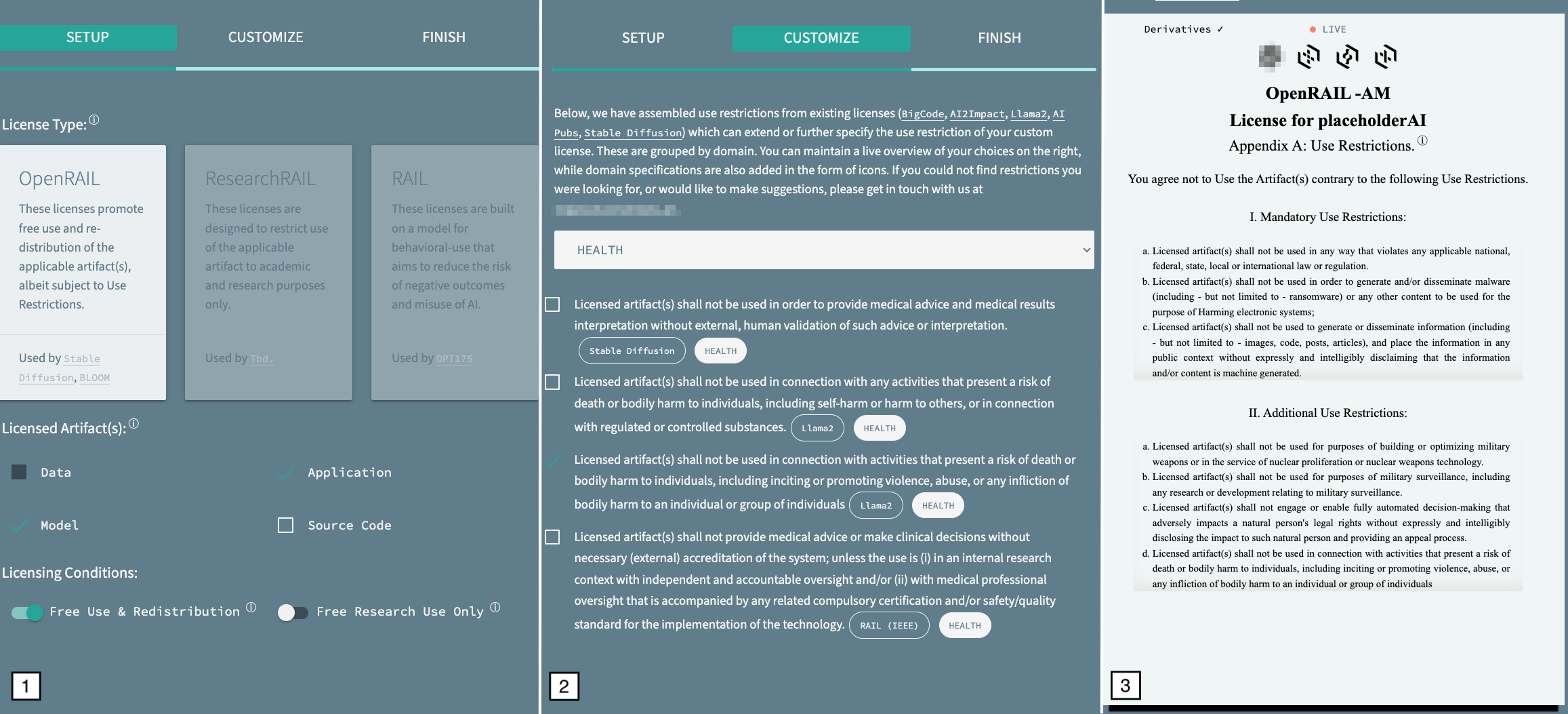}
    \caption{\textbf{Interface Designs for a Responsible AI License Generator.} \textit{1.} Setup panel for choosing a license for ``placeholderAI.'' On the left hand side, the type of artifact(s) specification and derivation handling can be specified by the user.
    \textit{2.} Customization panel for adding further use restrictions. On the left hand side, a dropdown menu allows selection of use restrictions by domain. 
    \textit{3.} On the right hand side, ticked selections are added to the preview in a separate block at the bottom. Note that icons related to the domains of selected restrictions are added to the top row.}
    \label{fig:interfaceDesign}
\end{figure*}

\textbf{Standardization of Responsible AI Licenses is Necessary.} Based on our analysis we take the position that the proilferation of responsible AI licensing is good, as it provides options to developers who are worried about misuse. However, there is a need for standardization to avoid confusing users or diluting impact while enabling some customization. We believe that there are tools that could support customizing licenses that still satisfy a set of core criteria. This is pertinent given that AI developers who aim to mitigate harmful uses of the artifacts they share will need to tailor licenses for the specific domain, context of use, and type of AI system. Below, we gather implications for license tooling development, considering how possible tools may support people (most likely without exhaustive legal expertise) in their practice of licensing the AI artifacts they are creating.

\textbf{A Community-Oriented License Generator.} 
An actionable route to standardized and customizable licenses is a generator purpose-built for AI licenses. Existing RAIL licenses already follow a topology ameanable to structured generation (e.g., Open-/ vs. RAIL, sub-specifications for Application/Model/Source Code artifacts).\footnote{See e.g. \url{https://www.licenses.ai/blog/2022/8/18/naming-convention-of-responsible-ai-licenses}, accessed 12/01/2024.}  Such generators are an established means to allow people without commensurate legal experise to choose among potential licenses for their artifacts. For instance, the Creative Commons (CC) initative has published a tool\footnote{\url{https://chooser-beta.creativecommons.org/}, accessed 15/01/2023.} which takes users through the process of deciding how people may or may not use their creative work. In the case of AI licenses, standardized customization along use restrictions may arguably allow for more ML-specialized tailoring (as called for by the interviewees); which in turn places specific demands on a license generator. 

Responsible AI licenses, unlike CC-type creative works, are \textit{not} broadly artifact-agnostic: in practice it matters whether one is licensing a ready made application, a set of source code, or a model, andthere may be specific restrictions and conditions that need to apply to each. For instance, one could choose to release a model with behaviorial-use restrictions but release the accompying source code under an open source license. 
In terms of a license generator, a design priority would be the ability to specify the artifact(s) to be licensed (e.g., \textit{D}ata, \textit{A}pplication, \textit{M}odel, \textit{S}ource Code) with behavioral-use clauses.

Specifying use restrictions gives AI developers further means to appropriately customize licenses. However, the large number of possible use restrictions risks leading to many different licenses that could be generated could confuse 1) the licensor who may not be able to easily judge how restrictive a license should be, 2) the licensee who may find it difficult to know what the assets are, and are not, permitted to be used for. Thus, aggregating potential use restrictions by domain could help (e.g., privacy, disinformation, health).

We developed an interface design for such a license generator (see Fig.~\ref{fig:interfaceDesign}), that incorporates the above considerations while also providing a ``live'' preview of the RAIL-type license that is currently being generated. The protoype shown includes a  base set of mandatory use-restrictions, that were selected based on our analysis of existing licenses. Icons related to the domains of selected restrictions
are added to the top of the license being generated.

\textbf {Code Scans for AI Licenses.} Similar to a license generator, it is feasible to construct tools which act as aids for AI developers when sharing and maintaining their created artifacts on platforms such as Huggingface and GitHub. An open question as far as licenses are concerned is the compability of dependencies (e.g., other repositories, third-party code, etc.) with the requirements of a responsible AI license. When dependencies exist that are licensed under a GPL license, for instance, a developer may not be able to select an OpenRAIL license for their artifacts due to requirements imposed by the GPL license. Alternatively, even when there are dependencies licensed under different responsible AI licenses, the arising artifact, will most likely need to be compliant with the stricter set of usage restrictions. 

In practice, AI systems---just like any contemporary software system---may integrate many dependencies, complicating this process further. To this end, a tool that scans existing dependencies for their licenses may aid AI developers in figuring out whether and which clauses are actually applicable. Such a tool could crawl the repositories of a given platform, and be integrated, for instance, as a service in GitHub Marketplace or in a Huggingface Space; or be offered as a command line tool for local use. In either form, such support for the workflow of AI developers in the ways they manage and share their artifact(s) may encourage a more reflective licensing practice that reflects the actual constraints of given dependencies.

\section{Conclusion}

With the growth in large parameter AI models the need for tools to help responsibly release models has increased. The use of responsible AI licenses have grown substantially over the past two years, thanks to adoption from several high-profile model releases (e.g., BLOOM, LLaMA2, StarCoder, Stable Diffusion, FALCON). Open Source licensing has been influential thanks to the clear definition and standardization of licenses such as MIT and Apache 2.0. The proliferation of different responsible AI licenses risks becoming confusing and we take the position that standarized forms of customization, along with appropriate tooling could make them more user-friendly.

\section*{Impact Statement}

As is increasingly well understood by the community, machine learning and AI projects can have significant, broad, and uncertain impacts. The development of new forms of licensing provides researchers with the ability and resources to release assets while restricting uses that they anticipate will lead to negative consequences. Standardization of licenses does involve making some value judgements about what types of behaviors are higher risk than others; however, providing the ability for customization can add flexibility. For licenses directed at responsible use such as RAILs, this can be achieved practically by allowing for revision and extension of mandatory and additional use restrictions in individual license instantiations.

\bibliography{refs}
\bibliographystyle{plainnat}

\balance{}
\newpage
\appendix
\onecolumn

\section{Tree Map of Models by License}
\begin{figure*}[h]
    \centering
    \includegraphics[width=1\textwidth]{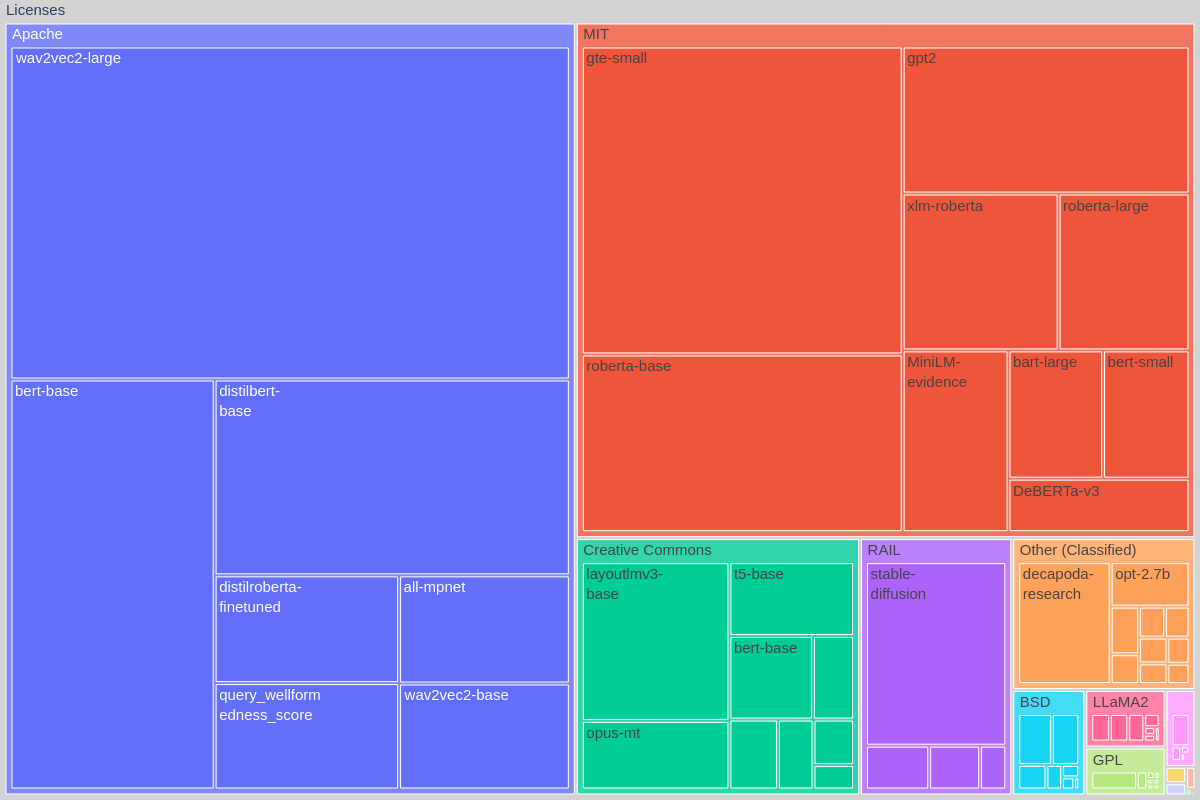}
    \caption{\textbf{Model Treemap by Downloads.} Treemap showing the top 10 models for each license weighed by the number of downloads}
    \label{fig:treemap}
\end{figure*}

\end{document}